# Substrate geometry affects population dynamics in a bacterial biofilm

Witold Postek[1,2], Klaudia Staskiewicz[1], Elin Lilja[1], Bartłomiej Wacław[1,3]

[1] Dioscuri Centre for Physics and Chemistry of Bacteria, Institute of Physical Chemistry, Polish Academy of Sciences, Kasprzaka 44/52 01-224 Warszawa, Poland

[2] Broad Institute of MIT and Harvard, 415 Main St, Cambridge, MA 02142, USA

[3] School of Physics and Astronomy, The University of Edinburgh, JCMB, Peter Guthrie Tait Road, Edinburgh, EH9 3FD, United Kingdom

Emails: wpostek@ichf.edu.pl; wpostek@broadinstitute.org; bwaclaw@ichf.edu.pl

**Abstract**
Biofilms inhabit a range of environments, such as dental plaques or soil micropores, often characterized by intricate, non-even surfaces. However, the impact of surface irregularities on the population dynamics of biofilms remains elusive as most biofilm experiments are conducted on flat surfaces. Here, we show that the shape of the surface on which a biofilm grows influences genetic drift and selection within the biofilm. We culture *E. coli* biofilms in micro-wells with an undulating bottom surface and observe the emergence of clonal sectors whose size corresponds to that of the undulations, despite no physical barrier separating different areas of the biofilm. The sectors are remarkably stable over time and do not invade each other; we attribute this stability to the characteristics of the velocity field within the growing biofilm, which hinders mixing and clonal expansion. A microscopically-detailed computer model fully reproduces these findings and highlights the role of mechanical (physical) interactions such as adhesion and friction in microbial evolution. The model also predicts clonal expansion to be severely limited even for clones with a significant growth advantage – a finding which we subsequently confirm experimentally using a mixture of antibiotic-sensitive and antibiotic-resistant mutants in the presence of sub-lethal concentrations of the antibiotic rifampicin. The strong suppression of selection contrasts sharply with the behavior seen in bacterial colonies on agar commonly used to study range expansion and evolution in biofilms. Our results show that biofilm population dynamics can be controlled by patterning the surface, and demonstrate how a better understanding of the physics of bacterial growth can pave the way for new strategies in steering microbial evolution.

**Introduction**
Bacterial biofilms are conglomerates of cells bound together by extracellular matrix containing compounds such as polysaccharides and nucleic acids (1). Found widely in natural ecosystems, biofilms play crucial roles in medicine (2, 3) and technology (4, 5). In all these contexts, the emergence of new genetic variants is a concern. For example, cells in a biofilm can acquire resistance to antibiotics through various mechanisms (6, 7), including *de novo* genetic mutations and horizontal gene transfer (8). Limiting the spread of such new variants is desirable, aligning with ongoing efforts to control biological evolution, which are gaining momentum (9).

The growth and population dynamics of biofilms are influenced by biochemical cues (10), competition (11), cooperation (12), cell death (13), and mechanical interactions (14, 15). In particular, the role of cell-cell and cell-surface interactions in the establishment of new variants has been explored in experimental (16, 17) and theoretical (18, 19) studies of colonies growing on agarose gel surfaces. In such colonies, bacteria primarily replicate at the expanding front due to nutrient depletion and waste accumulation at the center (20). When a new mutant arises at the front, it either "surfs" along the advancing front and forms a clonal cluster that expands into a new territory, or it gets outpaced by neighboring clonal populations and loses the competition for nutrients (16, 21-25). The probability of a new variant spreading depends on the interplay of mechanical interactions between bacterial cells and with the substrate, as well as the variant's fitness compared to the parent strain (13, 20, 26-33). These conclusions hold true also for three-dimensional, thick biofilms, which consist of a growing active layer and a quiescent bulk (34, 35).

A significant limitation in these studies has been the use of flat substrates. In natural environments, biofilms often grow on non-flat surfaces such as rocks and underwater stones (36), pipelines (4), mineralized surface of urinary catheters (37), pores of the human skin (38), porous beads in water treatment plants (39), marine snow (40), and colon crypts (41). Recent work on bacterial colonies growing on rough agar have shown that selection decreases and neutral drift increases compared to flat agar (42). A similar effect has been obtained by placing obstacles in the path of an expanding colony (43). Clonal dynamics is also affected in colonies encountering physical objects during expansion (44). However, these studies primarily focus on quasi-two-dimensional colonies that grow parallel to the rough surface, which does not consider the perpendicular growth of 3d biofilms. Consequently, there is a lack of experimental and *in silico* research on the influence of surface irregularities on the biofilm population dynamics.

To bridge this gap, we employ a microfluidics-based model system to investigate a more realistic scenario of a biofilm growing on a rough surface, where the top of the biofilm is sheared off by flow. We aim to understand how the population dynamics of genetic variants is influenced by substrate roughness. Specifically, we grow biofilms initiated with a mixture of fluorescently labelled cells in microscopic wells, featuring sine-like undulations of the bottom surface. We show that a moderate level of surface corrugation at the scale slightly larger than the cell size reduces both genetic drift and selection strength, enabling coexistence of clones with significantly different growth rates. This holds true even when the amplitude of substrate roughness is much smaller compared to biofilm thickness. We attribute this effect to cell-surface interactions that influence cell orientation and movement in the biofilms. The resulting velocity field hinders clonal mixing everywhere in the biofilm except close to the substrate, where bacteria can orientate and move parallel to the surface. Substrate roughness strongly limits this movement, resulting in narrow clonal sectors that do not invade each other even when clones have different fitness. Our microfluidic experiments, coupled with mathematical modeling, underscore the role of mechanical interactions in biofilm growth and evolution, and suggest a simple and robust way of controlling biofilm genetic diversity by varying substrate roughness.

## Results
**Substrate roughness leads to clonal sectors mirroring the substrate geometry.**
We cultured biofilms using a microfluidic device (Fig. 1) comprising multiple micro-wells connected to a deeper main channel for delivering growth medium and bacteria to the wells, similar to previously published devices (45, 46). However, in contrast to those studies, each well (100x100x7 μm) had a sine-patterned bottom surface (Fig. 1A), with varying amplitude and period across different wells, including some with flat bottoms. The amplitude of surface height modulation was only a small fraction of the total well height, and the sine-like indentations did not physically isolate different regions of the biofilm. The quasi-two-dimensional well shape allowed for the formation of multi-layered biofilms similar to naturally occurring biofilms, but the biofilms remained thin enough for cell movement to be limited mostly to the XY plane. This facilitated imaging of the biofilm using a wide-field epi-fluorescent microscope.

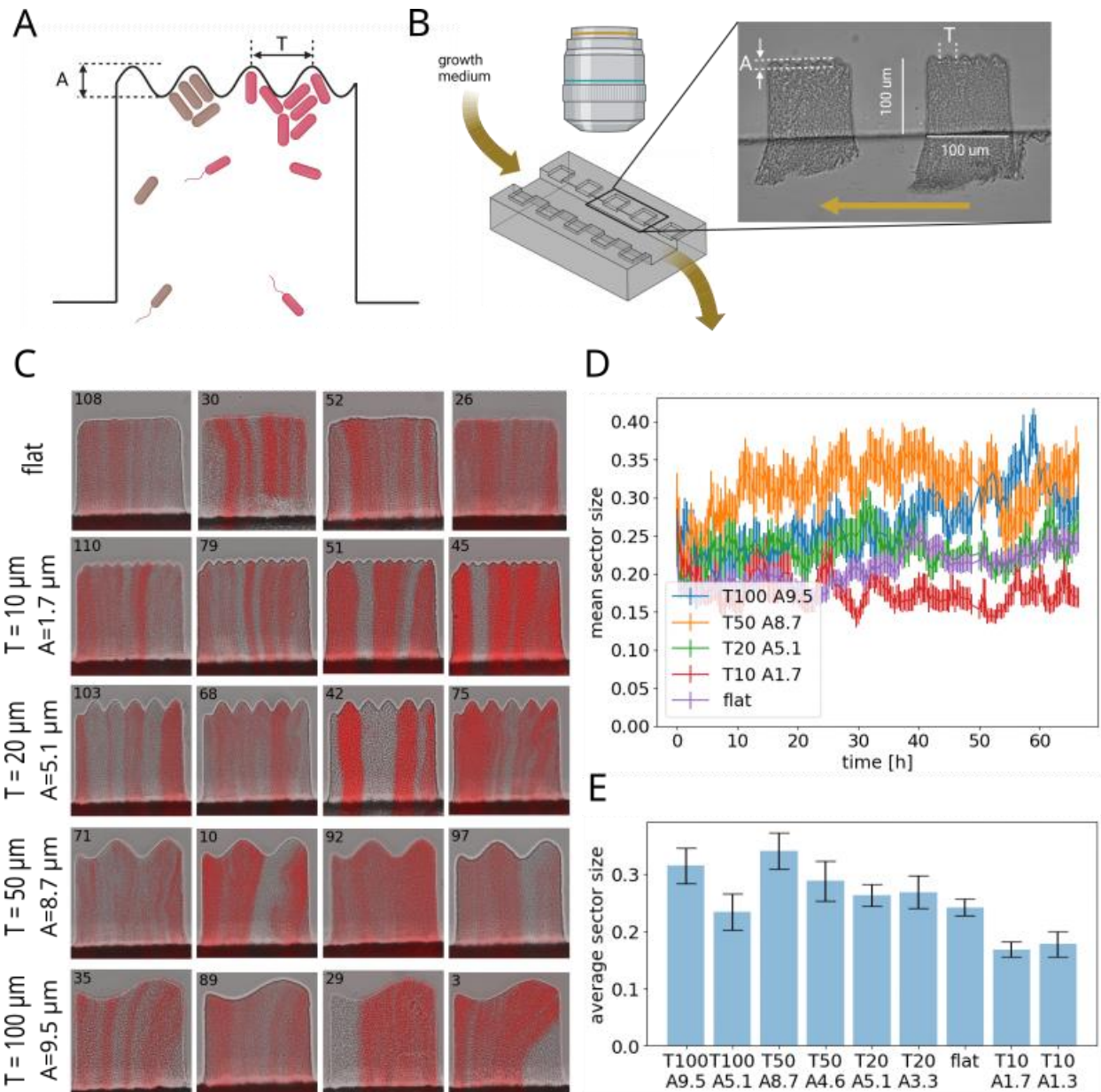

**Fig. 1. Substrate geometry affects the number and size of clonal sectors. (A)** Illustration of a single 100x100 µm well with a corrugated bottom for biofilm growth. **(B)** A simplified diagram of the experimental set-up. The actual microfluidic device has 240 wells with sine-like corrugations of 8 different amplitudes $A$ and periods $T$, each replicated 20 times, along with 80 flat-bottomed wells. **(C)** End-point snapshots of randomly selected wells with different configurations. The horizontal extension of quasi-vertical clonal sectors often matches the surface undulations. **(D)** Mean sector size versus time for five different combinations $(A, T)$. $t = 0$ h is 4 days post-inoculation. During the initial 4 days some biofilms fell out and were re-established. **(E)** Mean sector size as fraction of the well's width at the end point of live imaging (t=65 h, total growth time 6d 15h) for all well types (all combinations $(A, T)$). Error bars are S.E.M.

To examine the impact of growth on corrugated surfaces on clonal dynamics within the biofilm, we performed standing variation experiments using the biofilm-forming strain *E. coli* 83972 and its derivatives (*Methods*). We inoculated the microfluidic device with a 1:1 mixture of similarly-fit (*SI Methods*) red fluorescent (mKate) and non-fluorescent cells. Nutrient broth (LB) was pumped through the main channel, while we monitored all wells using fluorescence and phase-contrast microscopy. The experiment was conducted at room temperature (24-26°C) to reduce growth and clogging in the main channel. Nutrient flow trimmed excess biofilm extending into the main channel, thus allowing nutrients to diffuse into the wells. Biofilms quickly formed in all wells, with initial patches of fluorescent and non-fluorescent bacteria developing into sectors aligned with the direction of biofilm growth. The number of sectors and their positions matched the undulating pattern of the well bottom (Fig. 1C and SI Video 1), especially evident for undulations with periods much smaller than the well's width. Most sectors were parallel to the side walls, but some exhibited bending (e.g., biofilm no. 3 in Fig. 1C), suggesting non-uniformities in the velocity field within the well. Once established, the sectors maintained nearly constant width during live imaging (Fig. 1D).

To quantify the difference between sector patterns in various wells, we calculated the mean sector size as a fraction of the well's width (*Methods* and *SI Methods*). Figure 1D shows that long-period undulations result in larger sectors compared to flat-bottomed and short-period ones. To assess the impact of the amplitude $A$ alone, we compared wells with two different amplitudes but identical $T$. Figure 1E shows that reducing the amplitude has a lesser effect on the size of neutral sectors compared to decreasing the period $T$ for small $T$.

In contrast to similar experiments in bacterial colonies (16, 18, 21, 24, 25), the fluorescence within the sectors appeared more heterogeneous, indicating that the sectors were not entirely monoclonal. To understand how genetic diversity within each sector was reduced due to clonal expansion in the pockets of the undulating surface, we analyzed the inter-pocket diversity and compared it with intra-pocket diversity (SI Fig. S2). Specifically, we calculated the standard deviation of fluorescence intensity within each pocket (SI Fig. S2A). We then calculated the ratio $\rho$ of the average standard deviation of within-pocket fluorescence and the standard deviation of mean within-pocket fluorescence. SI Figure S2B shows that $\rho$ decreases with decreasing pocket size (decreasing $T$). Additionally, for a given $T$, the ratio $\rho$ is smaller for undulated surfaces than for flat-bottom wells when computed in "virtual pockets" positioned as they would be in the corrugated well. This indicates that while both fluorescent and non-fluorescent bacteria may be present in some pockets, undulations significantly hinder the mixing of adjacent bacterial populations.

**Bent sectors arise from the non-uniformity of the velocity field caused by heterogeneous adhesion.**
Some fluorescent sectors in Fig. 1C are bent, which suggest non-uniform flows in the biofilm (see also SI Video 1). This non-uniformity could be a result of differences in the growth rate or variations in adhesion to the surface in different parts of the biofilm. To distinguish between these two possibilities, we calculated the velocity field in the biofilm from short bright-field time-lapse videos recorded at a higher frame rate (1 fps) than those used for Fig. 1, using an optical-flow based method (*Methods* and *SI Methods*). We then calculated the local growth rate $g(x, y)$ using the continuity equation for an incompressible fluid with local mass generation: $g(x, y) = \vec{\nabla} \cdot \vec{v}(x, y)$. Figure 2A shows the velocity field superimposed on the field representing the growth rate for selected wells. It is evident that, except for a small gradient towards the main channel and the direction of growth (the vertical direction in the picture), the growth rate remains relatively uniform, even though the velocity field occasionally has a significant horizontal (perpendicular to the growth direction) component. This suggests that most of the observed heterogeneities of the velocity field are due to stronger adherence of parts of the biofilm to the sides of the wells.

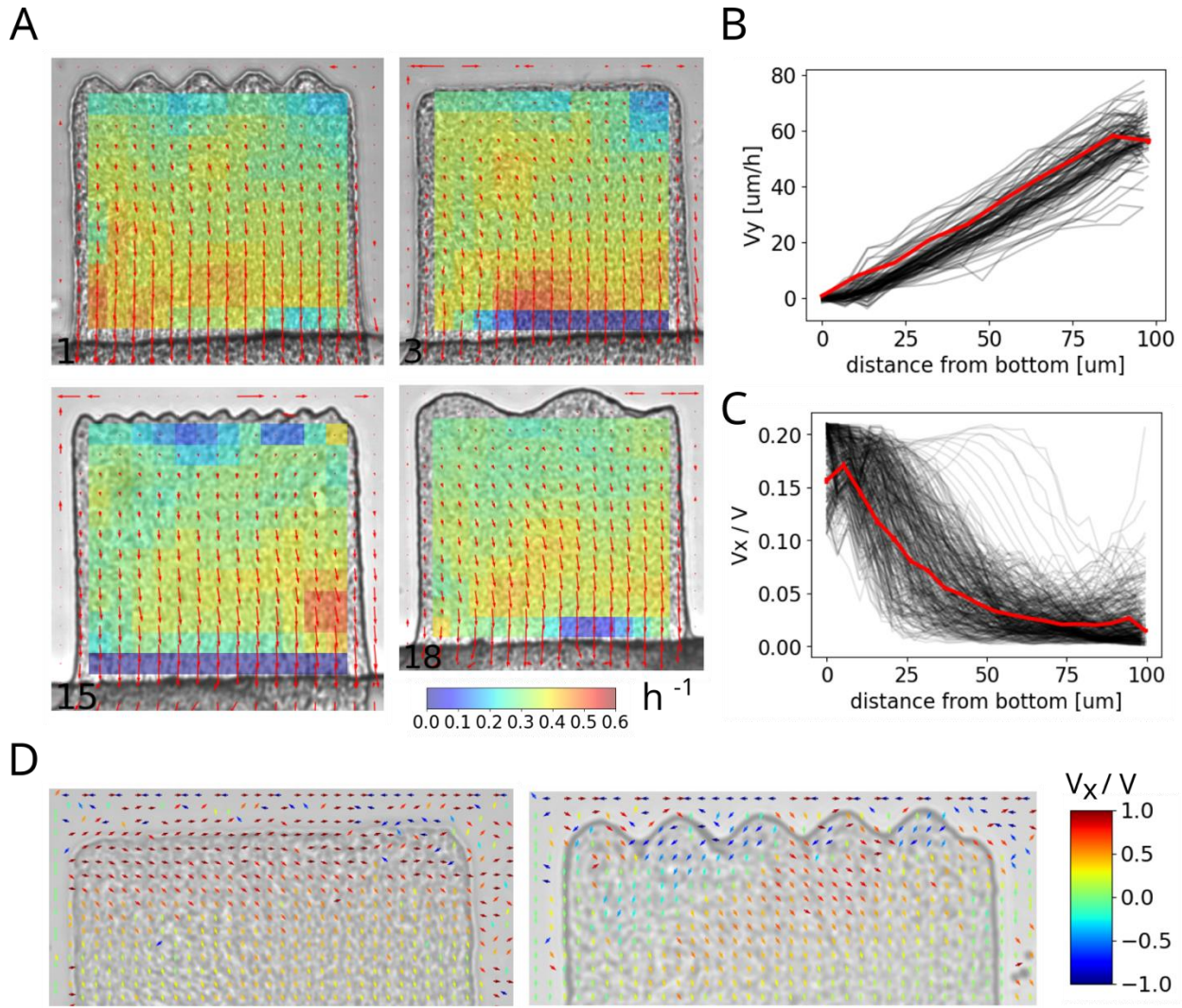

**Fig. 2. Velocity field in the wells. (A)** Examples of the velocity field (arrows) and growth rate (color map) for wells of different types. **(B)** Average vertical velocity (red curve) as a function of distance from the bottom surface of flat-bottomed wells. The individual black lines represent velocities measured at various horizontal positions (approx. 10 per well) within the well, and across different wells. **(C)** Average horizontal component of the velocity field (red curve) versus the distance from the well bottom. The black lines are as in panel C. **(D)** Zoomed-in plots of the velocity field near the bottom for a corrugated- and a flat-bottom well. Arrows are color-coded based on the $V_x/V$ component of the field (scale bar).

**Spatial isolation arises from the focusing property of the velocity field.**
Next, we investigated the properties of the velocity field that could explain the spatial isolation of the sectors. Figure 2B shows the vertical component of the velocity field, averaged across all horizontal positions and all flat-bottom wells. The linear increase of the vertical velocity with distance from the bottom indicates that the growth rate in our system is the same at all depths inside the biofilm. Such a linear velocity field facilitates the orientation of rod-shaped cells in the direction of growth (47). Figure 2C shows that the horizontal component of the velocity field, plotted as a fraction of the total velocity, decreases rapidly with distance from the bottom. Consequently, significant horizontal cell movement occurs primarily near the bottom surface of the well. This movement enhances genetic drift, promoting the establishment of a single clone in flat-bottom wells. To assess whether the presence of surface undulations hinders this movement, we examined the velocity field near the bottom of flat- and corrugated wells. Figure 2D shows that the horizontal movement of cells is indeed disrupted by the presence of surface undulations.

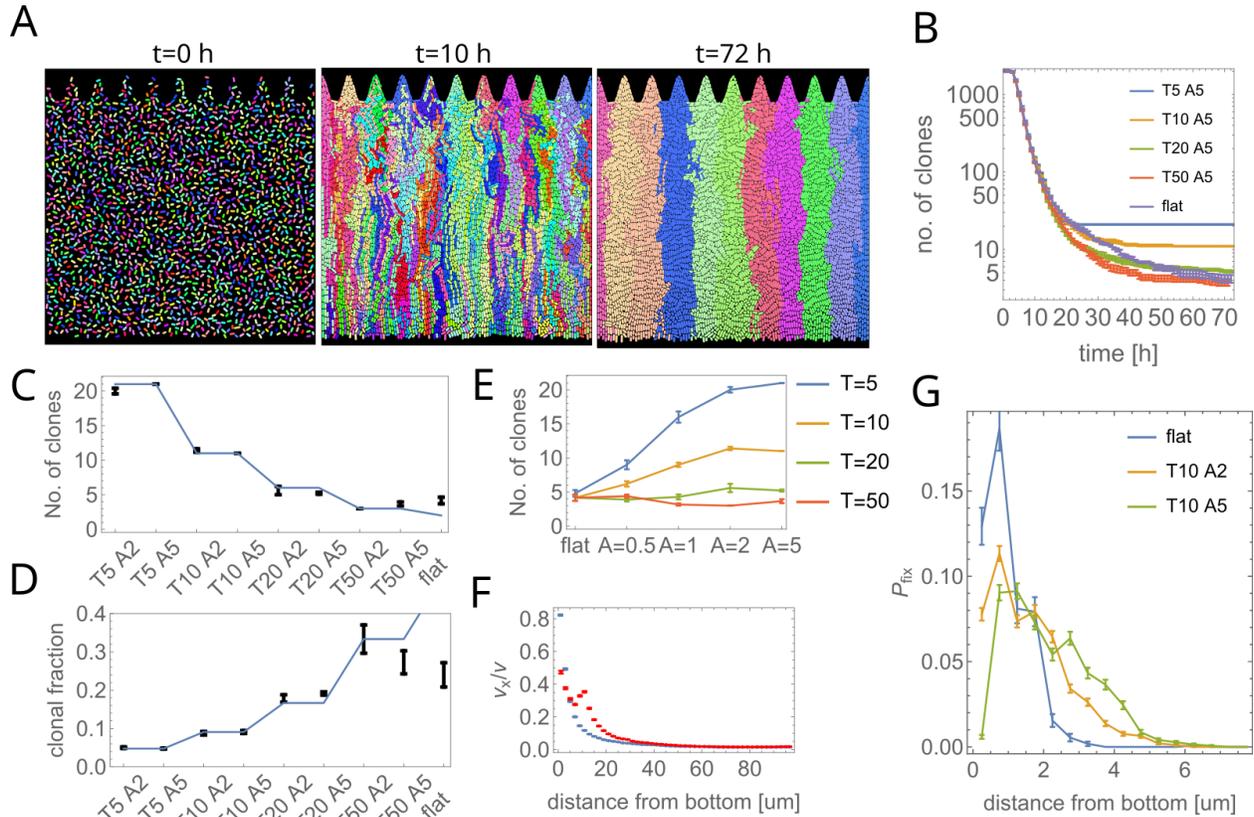

**Fig. 3. Computer model of the experiment**. **(A)** Simulation snapshots for $T = 10, A = 5$. **(B)** The number of clones as a function of time, for different $T, A$. **(C, D)** The number of clones after $t = 72$ h (C) and the mean fraction of the population occupied by a clone (D), for different $T, A$. The blue line shows theoretical average values for the number of clones $N_{\text{clones}} = (T + 100 \, \mu\text{m})/T$ and the fraction occupied by a single clone $f = 1/N_{\text{clones}}$ based the number of pockets, under the assumption of intra-pocket clonality. **(E)** The number of clones (as in panel C) as a function of amplitude $A$, for different periods $T$. **(F)** The lateral component of the local velocity field as a function of the distance from the bottom. The blue curve is for a flat bottom well, the red curve is for $T = 10, A = 5$. **(G)** Probability density function that the progeny of a cell, originated at $t = 2$ h at a distance $y$ from the bottom, survives until $t = 72$ h. In all panels, error bars = S.E.M.

**Computer model reproduces the experimental results.**
To understand the experimental results, we used a 2d computer model similar to the one reported before (20), to simulate a colony of rod-shaped bacteria in a square well with one open side, through which cells could escape and be removed from the system. We assumed that all bacteria in the biofilm divided at equal rates as supported by Fig. 2B. To facilitate tracking clonal sectors, the system was initialized with bacteria each assigned a different, heritable genotype, represented by a random color. The simulation ran for 72 h (a similar duration as the experiment) for different pairs of $A, T$. Figure 3A shows that, as expected, undulated bottom surface constrained the spread of genetic variants to sectors originating from within each sine-like pocket. The number of sectors decreased in time but eventually stabilized for corrugated surfaces (Fig. 3B). The final number of sectors closely aligned with the number of undulations (Fig. 3C, D), for a broad range of amplitudes $A$ (Fig. 3E). In contrast, a flat- or only slightly undulated bottom surface allowed bacteria to spread along it, leading to increased genetic drift. Figure 3F, which is analogous to the experimental figure 2C, shows that horizontal movement was indeed limited near the bottom surface. A new clonal variant could only establish if initiated by a cell close to the substrate on which the biofilm grows (Fig. 3G), otherwise the variant would be pushed out of the well by replicating cells that were closer to the substrate. This is the opposite of what occurs in bacterial colonies growing on a flat agar surface, where new variants can only establish if they emerge close to the colony edge (18, 20).

**Model prediction that undulations serve as suppressors of selection is confirmed experimentally.**
So far, we considered neutral variants, i.e., all genotypes (both in the experiment and in the model) had nearly identical growth rates. Since spatial patterning of the bottom surface decreases interactions between neighboring populations of bacteria, we reasoned that it should also prevent fitter variants from taking over. To test this hypothesis, we initialized our computer model with a 10:1 mixture of normal-growing: faster-growing clones with relative fitness $W = 1.5$ (*Methods*), and measured the fraction of the less-fit clone after $t = 72$ h. Figure 4A shows that the less-fit clone is quickly outcompeted by the faster-growing variant in flat-bottomed wells, but undulations impede its spreading, resulting in a substantial fraction (40-60%) of the less-fit clone remaining (Fig. 4B). Moreover, when simulated bacteria adhere to the surface of the well, the effect of selection is further diminished (dark green points in Fig. 4B). Therefore, our system acts as a suppressor of selection (48), allowing coexistence of diverse genetic variants.

To experimentally test this prediction, we inoculated our microfluidic device with a 1:10 mixture of mKate RIF[R] red-fluorescent, rifampicin (RIF) resistant strain and GFP RIF[S] green-fluorescent rifampicin sensitive strain. In the absence of RIF, the red strain had a small fitness disadvantage ($W_{R/G} \approx 0.9$) compared to the green strain (SI Fig. S5). By varying the RIF concentration, we could thus tune the selective advantage of the red versus green strain. After approx. 40 h of incubation in pure LB medium in the microfluidic device, which resulted in the establishment of fluorescent sectors, we changed the medium to 0.5 µg/ml RIF in LB, and then after another 50 h to 1 µg/ml RIF. Figure 4C shows that the fraction of the RIF-sensitive green fluorescent strain decreased in all wells during the exposure to RIF, with a more pronounced reduction in wells with flat bottoms, and corrugated bottoms with wider undulations. Subsequently, we replaced RIF with pure LB medium, leading to the successful re-establishment of the sensitive strain in corrugated wells with $T < 100$ µm (Fig. 4C, D, and SI Video 2). However, re-establishment occurred only rarely in flat-bottom wells or those with $T = 100$ µm. Figure 4E quantifies this effect by plotting the ratio of the sensitive strain fractions at two time points: after (t=182 h) and before (t=41 h) the exposure to RIF, for different well types. A similar trend is visible in the number of RIF-sensitive sectors (SI Figs. S3 ad S6).

The rate with which a fitter sector expands can be related to the relative fitness of the green versus the red strain (*SI Methods*). By fitting the model to the experimental time-series data for flat-bottom wells, we determined the relative fitness $W_{G/R} \approx 0.2$ (SI Fig. S4). We then ran the computer model without adhesion using initial clonal fractions obtained at the end of the 0.5 µg/ml RIF phase of the experiment from Fig. 4C, the relative fitness $W_{G/R} = 0.2$ and undulations amplitudes as in Fig. 4E. The model correctly reproduced the experimental results (Fig. 4F). Interestingly, repeating the same procedure for the model with adhesion sufficiently strong to affect the velocity field of the growing biofilm yielded a much worse fit (Fig. 4F). This does not mean that cells did not adhere to the walls but only that surface undulations were more important than adhesion for the outcome of our experiment.

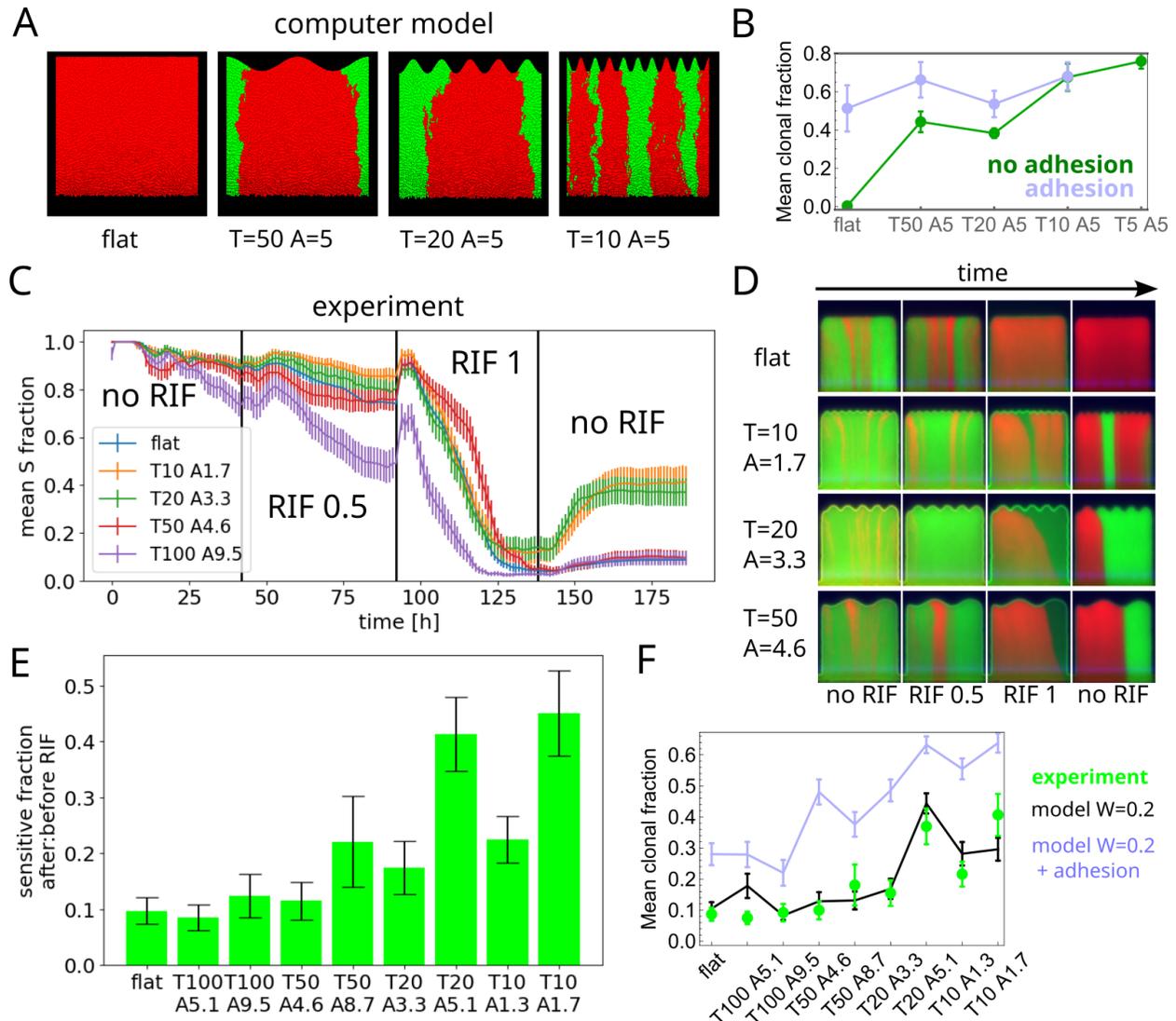

**Fig. 4. Corrugated surface limits selection. (A)** Computer simulation snapshots ($t = 72$ h) illustrate how corrugations constrain the spread of the fitter mutant (red, relative fitness $W_{R/G} = 1.5$). **(B)** The fraction of the less-fit green strain remaining in the biofilm after 72 h increases as the corrugation period $T$ decreases, for absent or weak adhesion. Strong adhesion nullifies this effect. **(C)** Experimental validation of the model: the fraction of RIF-sensitive (less fit) green strain as a function of time in an experiment in which the strength of selection was varied in time by adjusting the RIF concentration. The fitter red strain dominated in flat-bottomed and large-$T$ wells after transient RIF exposure, while rapid undulations (small $T$) significantly limited the spread of the fitter mutant. **(D)** Snapshots of wells corresponding to different phases of the experiment from panel C. **(E)** The ratio of the sensitive strain fractions at two time points: after ($t = 182$ h) and before ($t = 41$ h) the exposure to RIF. **(F)** The mean fraction of the sensitive strain in different types of wells (green points) is reproduced by the computer model without adhesion (black line). Error bars are S.E.M.

**Discussion**

We have shown that biofilm population dynamics depends on the shape of the surface on which the biofilm grows. By transitioning from a flat to a corrugated surface, we have demonstrated that surface undulations of appropriate period and amplitude can prevent clonal sub-populations from invading each other. This reduces genetic drift, enabling neutral variants to coexist, while also diminishing the impact of selection, preventing fitter variants from dominating over less-fit clones. These effects arise from the interplay of bacteria-surface interactions that influence cell orientation and the velocity field in the biofilm. Importantly, the depth of the undulations required to see these effects is only a small fraction of the biofilm's thickness, and the undulations do not physically separate different regions of the biofilm.

Our model system demonstrates significant differences compared to bacterial range expansion experiments on agar plates (21, 24, 25, 43, 49). Clonal sectors emerge despite uniform growth throughout the entire biofilm volume, whereas on agar plates a thin growing layer is necessary for clonal segregation to occur. Additionally, in our system, the roughness of the biofilm's top surface has no effect on its population dynamics, unlike in colonies on agar where the roughness of the leading edge significantly affects genetic drift (20). This is because in our system it is the growth at the bottom, not at the leading edge (top surface), that drives the dynamics of clonal sectors. Interestingly, "static" i.e., time-independent undulations of the bottom surface have the opposite effect compared to dynamically changing undulations of the expanding frontier of a growing colony (16, 43): genetic drift is reduced rather than enhanced by "static" roughness.

Computer simulations from Fig. 4 show that surface adhesion could have a profound effect on selection, reducing the influence of growth rate differences on the probability of establishment of fitter variants. Since in our model fitter clones must first expand close to the bottom surface to dominate the biofilm, we conclude that adhesion could partially counterbalance the effect of increased growth rate and hinder expansion. In our experiment, this effect does not seem to be very strong, perhaps because inter-cellular adhesion is stronger than adhesion to the walls. Nevertheless, we hypothesize that mutants with reduced adhesion could gain an additional, growth-independent selective advantage, similarly to what happens in bacterial colonies (50). However, the advantage might be short-lived: since adhesion is essential for biofilm establishment on surfaces, less-adherent variants could eventually cause the biofilm to detach. Consequently, conducting an experiment with a less adherent strain of bacteria would not be possible in our setup as it would result in biofilms falling out of the wells.

Surface adhesion is likely responsible for the observed persistence of neutral sectors in some flat-bottomed wells. In a well-mixed growing population whose size remains approximately constant due to the continuous removal of surplus population, one clone would eventually always reach fixation (51). However, in our system, adhesion may prevent fixation, particularly if certain cells (e.g., older ones) adhere stronger than others. Such cells may persist in the well for an extended period, leading to the formation of long-lasting sectors by their progeny.

A key mechanism driving clonal expansion on flat surfaces and within sine-like pockets is the horizontal component of the velocity field near the surface (Fig. 2). In the computer model (Fig. 3), this component arises due to "buckling", i.e., pressure-driven bending of chains of bacteria experiencing growth-induced compression (28, 30, 52, 53). This is supported by the lack of lateral motion of cells in simulations in which the height of the biofilm has been made much smaller, reducing mechanical stress and making cells more aligned with the flow, and the restoration of lateral movement upon increasing the friction coefficient (SI Videos 3, 4, 5). In the actual experiment, buckling manifests as bends in fluorescent sectors, particularly noticeable near the bottom. Notably, buckling is essential for the velocity flow to acquire a horizontal component, causing cells at the bottom of the well to orient themselves perpendicularly to the surface. For undulated surfaces, the onset of lateral buckling occurs earlier i.e., for smaller compression forces, because the curvature of such surfaces prevents cells from aligning into chains. This intuitive explanation clarifies why undulating surfaces hinder mixing: earlier buckling (more pronounced for surface undulations of higher amplitude) results in a faster transition from horizontal to vertical biofilm flow.

Since buckling is affected by adhesion and friction not only with the bottom and side walls but also the glass and PDMS surfaces, we expect that our results would be quantitatively – but not qualitatively – different, if we used taller wells ($h > 10$ μm) to accommodate more cell layers. Creating a fully three-dimensional biofilm would reduce the role of such boundary effects. Patterning the surface in two directions would be necessary to limit clonal expansion under such conditions.

We speculate that our findings generalize to naturally occurring biofilms that are relatively thin (a few hundred μm) to enable growth at the bottom. If the biofilm is thick enough to prohibit cell division at the bottom but its height remains limited due to mechanical shearing or flow, the shape of the surface it adheres to should become less important. Nevertheless, this might change in the presence of particles such as microplastics attaching to the biofilm and creating new adhesion points for cells.

Our results suggest that bacterial population dynamics in the biofilm can be controlled by patterning the surface. Many population genetics models (48, 54-58) rely on compartmentalization to influence population dynamics, which may not be applicable to growing biofilms. In contrast, our approach leverages the physics of the biofilm to achieve effective spatial separation. By manipulating the surface geometry, we can restrict the expansion of undesired genetic variants; we demonstrate this specifically for an antibiotic-resistant mutant. Sub-MIC transient antibiotic exposure is not uncommon and may lead to the emergence of resistance (46, 59). However, further research is required to see if surface patterning could be used for medical devices such as catheters or implants, which are prone to biofilm invasion (37). Similar strategies could also be utilized to stabilize engineered bacterial communities (60), in particular for the use in biosensing (61) or bioremediation, in which different bacterial ecotypes must often coexist together (62), and mutations in synthetic gene networks often present in such communities must be prohibited.

**Materials and Methods**

*Microfluidic device.* We fabricated a two-layer device made of PDMS attached to a glass slide (SI Fig. S1), with 240 micro-wells on both sides of a 500 μm-wide and 87 μm-deep channel. Each well measured 100x100x7 μm (width (X) x height (Y) x depth (Z)). The device contained 20 replicates of each of 8 sine wave/amplitude combinations, and 80 flat-bottomed wells. To fabricate the device, we utilized soft lithography, following well-established protocols (*SI Methods*). A photomask designed in AutoCAD (Autodesk) and printed by MicroLitho, UK, was used to expose a layer of negative photoresist on a silicon wafer. After developing, the negative mold was covered with PDMS (Sylgard, Dow Corning) mixed at a 1:10 ratio of curing agent to monomer and baked at 75°C for at least four hours. The resulting PDMS device was peeled off, oxygen-plasma treated, and bonded to a plasma-treated 1 mm-thick glass slide. After inserting PTFE tubing (Bola Bohlender, Germany, I.D. = 0.5 mm, O.D. = 1.0 mm), the device was connected to syringe pumps PHD2000 (Harvard Apparatus, USA) or (in some experiments) SyringeONE Programmable Syringe Pump (Darwin Microfluidics). We used plastic syringes with appropriate media (bacterial culture, LB, LB + rifampicin), depending on the type of experiment. Prior to use, all devices were flushed with 70% ethanol (Figs. 1-2) and 70% ethanol + 5% NaOH (Fig. 4) to sterilize the device, remove air and enhance bacterial adhesion.

*Bacterial strains.* We used the *E. coli* strain 83972 (63) and its fluorescent derivatives for all experiments, with either mKate (red) or GFP (green) being constitutively expressed from the bacterial chromosome. For the experiments with a RIF-resistant strain we used a variant of the red fluorescent strain with resistance conferred by a single point mutation of the *rpoB* gene. All strains easily adhered to surfaces and formed biofilms. The presence of the fluorescent reporter did not reduce fitness compared to the ancestral strain, but rifampicin resistance decreased fitness by about 10% (SI Fig. S5). All genetic modifications are detailed in *SI Methods*.

*Biofilm growth and imaging.* All experiments were performed at room temperature (24-26°C). We inoculated the microfluidic device through the attached PTFE tubing using dense bacterial cultures that were grown overnight in LB (Miller) broth (Carl Roth, Germany) at 37°C/180 rpm, diluted in fresh LB, mixed in desired ratios (1:1 or 1:10 as established by $OD_{600}$ measurements) and centrifuged to increase the cell concentration. After allowing bacteria to settle in the wells for approx. 30 min, we swapped the syringe with the bacteria to a new one filled with LB broth and initiated the flow. We used variable-flow rate protocols (*SI Methods*) to reduce clogging and biofilm growth in the main channel. For the experiment in Fig. 4, we replaced the LB medium with LB + RIF, and later switched back to LB, as described in the main text.

*Microscopy and image analysis*
Images were acquired on two fully automated Nikon Ti2-Eclipse epi-fluorescent microscopes equipped with automated XY stages, the Perfect Focus System, GFP and mCherry fluorescent filters, and controlled by MicroManager (64). Depending on the experiment, we used either 20x or 40x long-working distance Nikon objectives. Custom-written Python and Mathematica® code was used to load and process raw TIFF images outputted by MicroManager, and to analyze and plot the data. See *SI Methods* for details.

*Computer Model*
Computer simulations were conducted on the Edinburgh compute cluster at SoPA. We adapted the model previously reported (20), which involved representing bacteria as growing spherocylinders, with their movement confined to the XY plane, and interacting mechanically through Hertzian-like repulsion. The dynamics was overdamped with Stokes-like friction proportional to the velocity of the moving cell. We did not model nutrient diffusion and assumed that all cells of the same type grew at the same rate, regardless of their location in the well. We modelled repulsion from the walls of the well as a force that increased linearly with the overlap between the cell and the wall, with a stiffness constant $10^7$ pN/µm sufficiently large to prevent the overlap to increase beyond a fraction of a µm. Adhesion between bacteria and walls was represented using elastic springs (spring constant $k = 10^6$ pN/µm), connecting the center line of the cell to the closest point of first contact on the wall. The springs broke when extended by more than 20% (Fig. 4B) and 5% (Fig. 4F) of their initial length.

**Data, materials, and software availability**
All code (C++, Jupyter notebooks, Mathematica notebooks), processed image data and simulation results are available at https://github.com/Dioscuri-Centre/biofilms_on_corrugated_surfaces . Due to large file sizes (several TBs), raw image data have not been uploaded. Access to such data will be provided on request.

**Acknowledgements**
This work was supported by the following grants: the Bekker Programme of the Polish National Agency for Academic Exchange (NAWA) grant no. PPN/BEK/2020/1/00333/U/00001 (W.P.), the START scholarship of the Foundation for Polish Science (FNP) grant no. 069.2021 (W.P), NAWA Polish Returns grant no. PPN/PPO/2019/1/ 00030 /U/0001 (W.P. and B.W.), and Dioscuri, a programme initiated by the Max Planck Society, jointly managed with the National Science Centre in Poland, and mutually funded by Polish Ministry of Science and Higher Education and German Federal Ministry of Education and Research, grant no. UMO-2019/02/H/NZ6/00003 (W.P., K. S., E. L., B. W.). Figure 1 was created with BioRender.com.

**Competing interests**
The authors do not report any competing interests.

**Use of AI tools**
ChatGPT was used to refine the text and improve its flow, although no ChatGPT-generated sentences were copied verbatim into the manuscript. The manuscript was read and approved by all authors.

# Supplementary Information

Substrate geometry affects population dynamics in a bacterial biofilm.

Witold Postek, Klaudia Staskiewicz, Elin Lilja, Bartłomiej Wacław

## Supplementary methods

*Microfluidic device architecture*
The microfluidic chip used in our study was a two-layer device consisting of a PDMS mold attached to a 1 mm-thick glass slide (SI Fig. S1). The device featured one inlet and two outlets equipped with pillar-based filters to prevent inflowing debris. The main channel was 500 µm wide. Micro-wells for culturing biofilms extended perpendicularly from the main channel; each was 100 µm wide (parallel to the main channel) and 100 µm deep (perpendicularly to the long axis of the main channel). The final height of the main channels and the wells obtained through soft lithography (see below) was 87 and 7 µm, respectively, so that only a few layers of cells could fit into each well, whereas the much taller main channel allowed for rapid nutrient medium flow. The 7 um well thickness facilitated optical imaging while at the same time ensuring that most bacteria interacted with other bacteria in the bulk of the biofilm rather than with the top (PDMS) or bottom (glass) surfaces. The biofilms were continuously trimmed by the flow in the main channel, ensuring a steady supply of nutrients to the deepest layers of the biofilm.

Each microfluidic device had 240 micro-wells, with 120 on each side of the main channel. The bottom of the well, which faced away from the main channel, was designed to be either flat or undulated. The shape of the undulations was a sine function of different periods and amplitudes (all dimensions in µm):
$(T, A) = ((100, 9.5), (100, 5.1), (50, 8.7), (50, 4.6), (20, 5.1), (20, 3.3), (10, 1.7), (10, 1.3))$. Due to the limitations of soft lithography and mask resolution, it was not possible to have the same amplitude $A$ for all periods $T$. The reported amplitudes are actual amplitudes obtained from microphotographs of the device. Each device contained 20 replicates of each sine wave / amplitude combination, as well as 80 flat-bottomed wells. The CAD design of the device is available on GitHub (1).

*Soft lithography*
To create a negative of the microfluidic device, we followed well-established photolithography protocols (2). First, we designed a photomask in AutoCAD (AutoDesk) and had this mask printed by an external company (MicroLitho, UK). We then covered a 3-inch silicon wafer (Microchemicals) with an SU-8 photoresist (Kayaku Advanced Materials) using a spin coater (Laurell, USA). After a soft bake on a programmable hot plate (4 minutes at 95°C), we exposed the wafer through the photomask representing the micro-wells layer, using a MJB4 mask aligner (SÜSS MicroTec). The second layer of SU-8 was then spun on the wafer, and the wafer was again soft-baked (5 minutes 65°C, 20 minutes 95°C). Edge bead-removal procedure was applied using the spin coater by covering the edge of the spinning wafer with photoresist developer mr-600 (Micro Resist Technology, Germany), deposited through a syringe, to unravel the alignment marks from the first layer. The wafer then underwent another soft bake (5 minutes 65°C, 20 minutes 95°C). Next, the wafer was illuminated through the photomask representing the main channel, and a post-exposure bake (5 minutes 65°C, 10 minutes 95°C) was performed. The specific spinning times, speeds, soft bake/hard bake times, and illumination parameters were obtained from the SU-8 manufacturer's (Kayaku) protocols. The wafer was developed with mr-600 developer (Micro Resist Technology, Germany) according to the manufacturer's protocol. After the development, the wafer was hard baked at 150°C for 30 minutes.

*Microfluidic device casting*
The wafer with the photo-resist negative of the device was covered with PDMS (Sylgard, Dow Corning) mixed at a 1:10 ratio of curing agent to monomer, and baked at 75°C for at least four hours. The cured PDMS was peeled off the wafer and inlet and outlet holes were created using a 1 mm-diameter biopsy puncher (Kai Medical). The PDMS device was activated with oxygen plasma alongside a glass slide in a plasma cleaner (Harrick Plasma, USA) for 60 seconds. Following this, the PMDS mold was gently placed on the glass slide and lightly pressed with a pair of metal tweezers to ensure proper bonding of the PDMS to the glass.

*Bacterial strains*
*E. coli* 83972 (DSM number 103539) was obtained from DSMZ GmbH. The red fluorescent mKate marker (under the control of the constitutive promoter *PtetO1*) was introduced into 83972 using plasmid mediated gene replacement (3, 4) replacing the *galK* gene. The strain used to amplify the mKate marker with the *PtetO1* promoter from was a gift from Meriem El Karoui (5). The green fluorescent GFP marker under the control of the constitutive *PA1* promoter was introduced into *E.coli* 83972 using plasmid mediated gene replacement, replacing the *galK* gene. The GFP marker with the *PA1* promoter was amplified from plasmid pGRG36-Kn_PA1-GFP(6) Plasmid pGRG36-Kn-PA1-GFP was a gift from Frank Rosenzweig (Addgene plasmid # 79088 ; http://n2t.net/addgene:79088 ; RRID:Addgene_79088). The primers used to amplify the upstream and downstream regions surrounding the *galK* gene of 83972, as well as the primers used to amplify the mKate and the GFP markers can be found in Table S1. Crossover PCR was used to anneal the 83972 homologous regions with the mKate and GFP markers, and these constructs were then inserted by restriction digestion and ligation into the plasmid pTOF24 (4) used for the gene replacement.

A rifampicin (RIF) resistant version of 83972 with the mKate marker was generated by plating the strain on LB agar plates supplemented with 100 µg/ml rifampicin, and randomly picking a resistant colony.

| | |
|---|---|
| 1.galK_up_fwd | AAA AAC TGC AGA CAC TGG TTA GCC GTT GTA C |
| 2.galK_up_rev | ATA GGG ACT CGA TTTC TTA CAC TCC GCA TTC |
| 3.mKate_gal_fwd | GGA GTG TAA GAA TCG AGT CCC TAT CAG TGA |
| 4.mKate_gal_rev | CGG GAG TTT CGT TTA TCT GTG CCC CAG TTT |
| 5.galK_down_fwd | GGG CAC AGA TAA ACG AAA CTC CCG CAC TGG |
| 6.galK_down_rev | AAA AAG TCG ACT GAT CGC CAT CAT CTG AAC T |
| 7.Gal_up_fwd | AAA AAC TGC AGT GAC GAT CGT TCT GGT TCA C |
| 8.Gal_PA1_up_rev | TGA TAA CCG CTA CGG AAG AGC TGG TGC CTG |
| 9.PA1_gal_fwd | CCA GCT CTT CCG TAG CGG TTA TCA AAA AGA |
| 10.GFPt7_gal_rev | GGA GTG TAA GAA TCA GCA AAA AAC CCC TCA |
| 11.Gal_GFPt7_down_fwd | GTT TTT TGC TGA TTC TTA CAC TCC GGA TTC |
| 12.Gal_down_rev | AAA AAG TCG ACA CAC TGG TTA GCC GTT GTA C |

Table S1. Primers used to amplify the upstream (1,2) and downstream (5,6) regions of the galK gene of 83972, with overlap to PtetO1_mKate, primers used to amplify mKate with the PtetO1 promoter (3,4), primers to amplify the upstream (7,8) and downstream (11,12) regions of the galK gene of 83972, with overlap to PA1_GFP, and primers to amplify GFP with the PA1 promoter (9,10).

*Bacterial cultures*
Single colonies were grown from frozen stocks on Luria broth agar plates at 37°C for 24h. Liquid cultures were prepared by inoculating 2 mL LB (Miller) broth (Carl Roth, Germany) using a single colony, and incubated overnight in a shaken incubator (37°C at 180 rpm). After the overnight incubation, the cultures were diluted in 10 mL of fresh LB (Miller) and further incubated at 37°C for at least 6h. The cultures were then mixed in a desired proportion (approx. 1:1) according to their optical densities (OD600), with the exception of the experiment in Fig. 4, for which 10 ml overnight cultures were first centrifuged at 4000 rpm for 2 min, re-suspended in 1 mL to concentrate them 10x, and finally mixed in a 1:10 ratio (GFP:RIF$^S$ to mKate:RIF$^R$).

*Growth media and flow control*
We used LB Broth (Miller) sterilized by autoclaving at 121°C for 15 min, and supplemented with rifampicin (RIF) (Merck KGaA, Germany, concentrations as in the main text) for the experiments in Fig. 4. The medium was delivered by syringe pumps using plastic syringes (BD, USA). The syringes were connected to the microfluidic devices with PTFE tubing (Bola Bohlender, Germany, I.D. = 0.5 mm, O.D. = 1.0 mm), with identical lengths for both outlets to ensure equal hydraulic resistance. 0.5 mm OD needles were used to connect syringes to the tubing. Syringe pump PHD2000 (Harvard Apparatus, USA) was used for experiments in Figs. 1-2. For Fig. 4 we used a SyringeONE Programmable Syringe Pump (Darwin Microfluidics).

*Biofilm experiments*
For experiments in Figs. 1-2, the microfluidic device was flushed with 70% ethanol using a syringe mounted on the syringe pump, and left for 10 minutes. The syringe was then replaced with a syringe filled with LB medium and the microfluidic device was flushed with LB. Next, the syringe was replaced with a syringe filled with a bacterial suspension. The suspension was pumped through the device and left at room temperature for 30 minutes. Finally, the syringe was replaced with an LB medium syringe, and the device was placed on the XY microscope stage for imaging. The flow rate was set to 50 µl/h for the initial 16 h and changed to 200 µl/h afterwards.

For experiments in Fig. 4, the microfluidic device was prepared by first flushing it with a solution of 5% sodium hydroxide (Carl Roth, Germany) in 70% ethanol using a syringe mounted in a syringe pump, at a rate of 2 ml/h for 8 min. Afterwards, the device was flushed with 70% ethanol, followed by LB medium. A dense bacterial suspension was then introduced into the device using a syringe pump at a flow rate of 3 ml/h. When bacteria showed up in the main channel, the flow rate was reduced to 1 mL/h and periodically turned on and off for about 30 min, encouraging the bacteria to attach, while continuously imaging the wells until they contained hundreds of cells/well. The bacterial syringe was then replaced with an LB syringe. The flow rate was set to 50 µL/h for 500 µL (10 h), followed by an alternating fast/slow flow of 3 ml/h for 10 µL and 50 µl/h for 16 µL to reduce clogging of the main channel in the 180h-long experiment. Throughout the experiment, the medium was replaced with LB+RIF at 0.5 ug/ml, LB+RIF at 1 ug/ml, and pure LB as described in the main text.

We ran all experiments at room temperature (24-26°C); this helped to limit biofilm growth in the main channel.

*Microscopy*
Images were acquired using two fully automated Nikon Eclipse Ti2-E epi-fluorescent microscopes with automated XY stages and the Perfect Focus System, and controlled by MicroManager (7). One microscope used an ORCA-spark Digital CMOS camera C11440-36U (Hamamatsu, Japan), the other one an Andor Zyla 4.2 SCMOS camera (Oxford Instruments, UK). Depending on the experiment, we used two different objectives (20x and 40x). To acquire fluorescent images of mKate and GFP strains, we used filters with excitation/detection wavelengths of 532 – 554 nm/ 573 – 613 nm, and 457.5 – 487.5 nm/ 502.5 – 537.5 nm.

*Image and data analysis*
To analyze the data, we utilized custom Python and Mathematica® code to load and process the TIFF images generated by MicroManager. The code allowed us to perform the necessary data analysis and create plots for visualization. The code is available on GitHub as Jupyter and Mathematica notebooks (1).

*Figure 1.* To obtain the mean sector size, we calculated the autocorrelation function of pixel brightness along the cut through the middle of the well at ½ of the distance from the opening to the bottom surface. We then took the position of the first minimum of the autocorrelation function and used it to represent the mean sector size. This method offered several advantages compared to counting sectors in a thresholded image: (i) it was not sensitive to the absolute value of fluorescence signal, which could differ between wells and changed over time due to variations in RFP expression, (ii) it was not sensitive to minor variations in pixel intensity for neighboring pixels, thus avoiding sector over-counting caused by very narrow darker or lighter "streaks" within larger sectors. Moreover, the algorithm (iii) correctly predicted the average size of stripes for synthetic data with interleaved bright and dark stripes, (iv) it offered a more objective approach than manual sector counting.

*Figure 2.* We imaged the biofilm every second for about 3 minutes using the 40x objective. The biofilm grew only minimally during this time, and individual cells moved less than a pixel per time frame. Inspired by the methods shown before (8, 9), we determined the velocity field $\vec{v}(x,y)$ from subtle changes in pixel brightness caused by the biofilm's local motion. This method did not require tracer particles or feature detection within the biofilm. Briefly, we assumed that the flow in the biofilm caused the pixel intensity field $I(x,y,t)$ to evolve during a small time interval $dt$ as follows: $I(x,y,t+dt) = I(x,y,t) + dt\, \vec{v}(x,y) \cdot \vec{\nabla} I(x,y,t)$, where $\vec{\nabla}$ represents the (discrete) gradient operator. This linear set of equations (one for each pixel) could be solved numerically for $\vec{v}(x,y)$ using pixel intensities at two time points. Since the system of equations was underdetermined, we binned the image 32x in each direction and solved for the two components of the average velocity field within each 32x32 block of pixels using the least squares method. Additionally, we automatically selected the time separation $dt$ for each bin that yields the most accurate estimate for $v(x,y)$ at that location without violating the assumption of small changes ($v\, dt < 0.1$ bin size).

*Figure 4.* We first determined the intensity profiles in red (mKate) and green (GFP) channels along a horizontal line cutting through the well at ½ of its height, for all time points and wells. Pixel intensities in both channels were rescaled by the minimum and maximum values from the entire acquisition (all wells and time points), and then logarithmized as follows: $I_{\text{rescaled}} = \ln(I_{\text{raw}} - I_{\min} + 1) / \ln(I_{\max} - I_{\min})$. The 1d array obtained in this way was convolved with a top hat function of width 10 pixels to reduce noise. Next, we detected the boundaries between red and green sectors based on pixel intensity in each channel. A pixel was classified as belonging to a "red" sector if its intensity was higher in the red channel than in the green channel, and vice-versa for the "green" sector. Using sectors boundaries determined in this way, we calculated the number of sectors and their sizes for each well and time point. To find the fraction of sensitive (green) strain, we calculated the proportion of "green" pixels ($I_{\text{G,rescaled}} > I_{\text{R,rescaled}}$) along the line of pixels. This method worked reliably once green and red bacteria separated into sectors, which is the reason for the apparent decrease of the mean sensitive fraction in Fig. 4C during the first part of the experiment (no RIF), despite the red strain having a small fitness disadvantage. We also confirmed that this method gave qualitatively similar results to the manual counting of sectors (Fig. S3).

*Doubling time of bacteria in the biofilm from Fig. 4*
We tracked the movement of easy-to-distinguish features (brighter spots and swirls) in fluorescence images of biofilms growing in flat-bottomed wells during the first phase of the experiment presented in Fig. 4 (pure LB, no RIF). If $y_t, y_{t+\Delta t}$ denote the distance of a feature from the bottom at times $t$ and $t + \Delta t$, then the growth rate $\alpha$ can be calculated as

$$\alpha = \frac{\ln\left(\frac{y_{t+\Delta t}}{y_t}\right)}{\Delta t}.$$

We calculated $\alpha$ for all flat-bottomed wells, using 2-3 traceable features per well. We obtained the average value $\alpha = 0.22 +/-0.01$ h$^{-1}$, corresponding to the doubling time ≈3 h. We did not observe any significant correlation between $\alpha$ and $y_t$ of the feature (Pearson correlation coefficient $r = 0.10$, p-value 0.40), which confirmed that growth did not depend on the depth (distance from the outlet) in the biofilm.

The growth rate was significantly lower in the biofilm from Fig. 4 than in the liquid culture (Fig. S5) at the same temperature. It was also lower than the growth rate from Fig. 2, in which the biofilm was less dense. Interestingly, the growth rate of single bacteria that occasionally attached to the surface of the main channel was very close to that of Fig. S5. Therefore, even if the growth medium flowing through the device was partially metabolised by bacteria living upstream, this was not significant enough to affect the growth rate.

In contrast to previous work (10), where mechanical confinement to a monolayer of cells has been suggested as a possible explanation of reduced growth rate, in our experiment the movement of cells is not constrained enough to cause axial compression, which could affect the growth rate (11). We hypothesize that cells in the biofilm switch to a slower-growing phenotype, perhaps as the effect of quorum-sensing (12).

*Relative fitness from sector expansion*

We used data from Fig. 4C to obtain the relative fitness $W_{G/R}$ of the sensitive green strain compared to the resistant red strain. We first selected flat-bottomed wells in which the initial fraction of resistant (red) cells was between 0.1 and 0.4 at the beginning of the 45 h-long 1 μg/ml RIF phase from Fig. 4C. The [0.1, 0.4] range was chosen because the sector detection algorithm performed well in this range. Figure S4A shows plots of the resistant fraction versus time for the selected wells. We then ran the computer model in flat-bottomed wells for different resistant initial fractions and different relative fitness values $W_{G/R}$. We assumed the doubling time of 3 h (see the preceding section) and the same duration of the simulation as the 1 μg/ml RIF phase. We selected runs for which the initial fraction was within $\pm 0.01$ of the experimentally determined values, thus the simulated curves (Fig. S4B) had the same initial distribution of resistant fractions as the experimental curves. Finally, we compared the theoretical and simulated average resistant fraction versus time curves obtained for different $W_{G/R}$, and determined that the best fit was given by the model assuming $W_{G/R} = 0.2$ and no adhesion.

*Relative fitness from liquid culture growth*

We used the method from previous work (13) to determine the growth rate of each strain at different concentrations of RIF. Briefly, we incubated bacteria in LB with different concentrations of RIF (0, 1, 1.5, 2, 2.5, and 3 μg/ml) in a 96 well micro-plate (200 μl/well) inside a plate reader, starting from two different initial cell densities, $N_0$ in rows A-D and $N_0/10$ in rows E-H. The $N_0$ dilution was an exponentially growing culture of OD600 $\approx 0.1$. We used one column of the 96 well plate for each RIF concentration. The plate reader (BMG LABTECH FLUOstar Optima) was set to incubate the plate at 25 °C in a cold room (4 °C) for enhanced temperature stability, with orbital shaking at 200 rpm for 10 s prior to OD measurement.

We measured the optical density (OD) of each culture every 2 min to obtain growth curves for approx. two days of incubation time. The exponential growth rate was then determined from the time shift between the growth curves for initial bacterial concentrations $N_0$ and $N_0/10$.

Figure S5 shows that the two fluorescently-marked strains have almost identical growth rates in the absence of RIF. The sensitive green strain grows about 10% faster without RIF ($W_{G/R} = 1.1$), but at 1 μg/ml used in experiments from Fig. 4 its relative fitness is $W_{G/R} \approx 0.6$ compared to the red resistant strain. This differs from the estimate from the previous section based on fitting the computer model to the sector expansion experiment in the biofilm. We attribute this to a slower growth rate in the biofilm than in the liquid culture, which makes the RIF$^S$ strain more sensitive to RIF in the biofilm. Indeed, slower growth correlates with increased susceptibility: the minimum inhibitory concentration of RIF for the sensitive GFP strain is 16 μg/ml at 37 °C (fastest growth), 8 μg/ml at 30 °C and 2 μg/ml at room temperature (slowest growth).

*Relative fitness from a competition assay in bulk cultures*

The relative fitness of the *mKate*-labelled 83972 strain as compared to the wild type was determined by competing the strains against each other in LB broth, shaking at room temperature (25 °C). Overnight cultures of both strains were mixed in a 1:1 ratio, diluted 1000-fold into four replicate 10 mL cultures, and incubated until they reached the stationary phase. The ratio of the strains before and after competition was determined by plating $10^6$ dilutions of the initial mixture and of each of the end cultures on LB agar, and the number of red fluorescent and non-fluorescent CFUs was counted. The relative fitness (W) of the *mKate* strain was then calculated by the following formula (14), where $R_0$ is the frequency of *mKate* before the competition, $R_1$ is the frequency of *mKate* after the competition, and $F$ is the fold increase of bacteria during the competition as determined by the dilution:

$$W = \frac{\ln\left(\frac{R_1 \times F}{R_0}\right)}{\ln\left(\frac{(1-R_1) \times F}{1-R_0}\right)}.$$

We obtained $W = 0.97 \pm 0.01$, i.e., a very small fitness disadvantage of the mKate-labelled strain. For the purpose of our biofilm experiment, in which selection is very strongly suppressed as explained in the main text, we can consider both variants to have the same fitness.

## Supplementary Figures

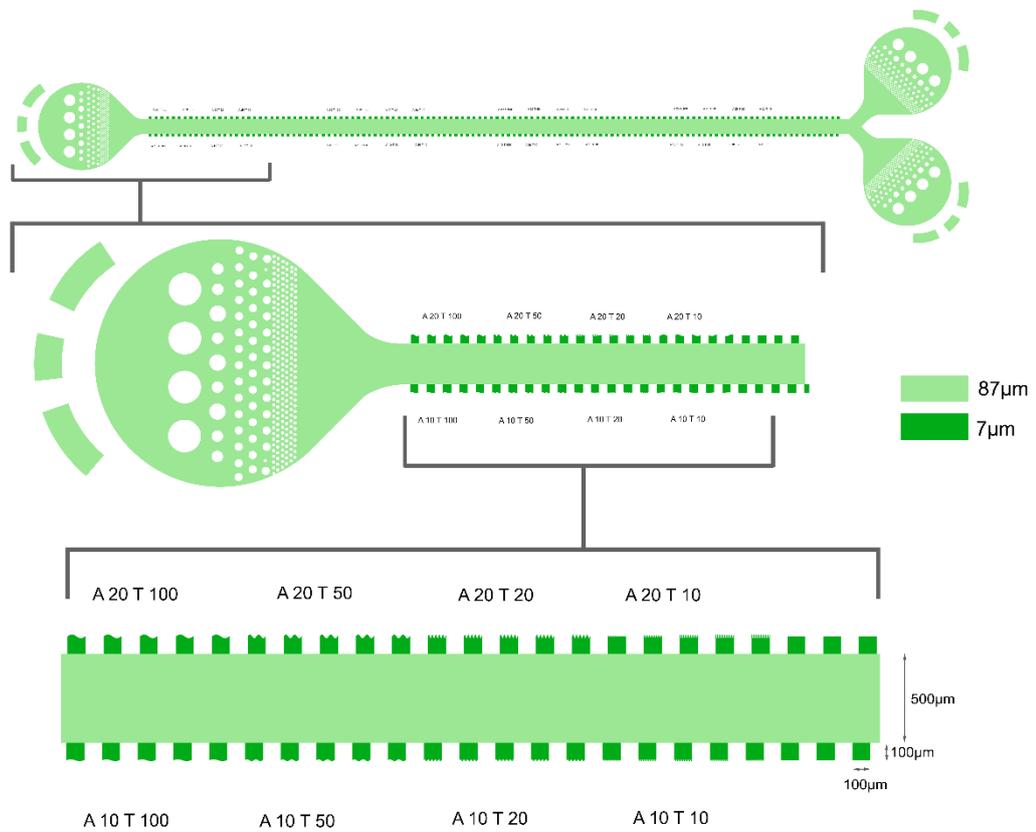

**Fig. S1. Design of the microfluidics device.** Top panel: the entire device. Middle and bottom panels: small sections of the main channel (width = 500 μm) with 100x100 μm wells of different types (corrugation period and amplitude) visible on both sides.

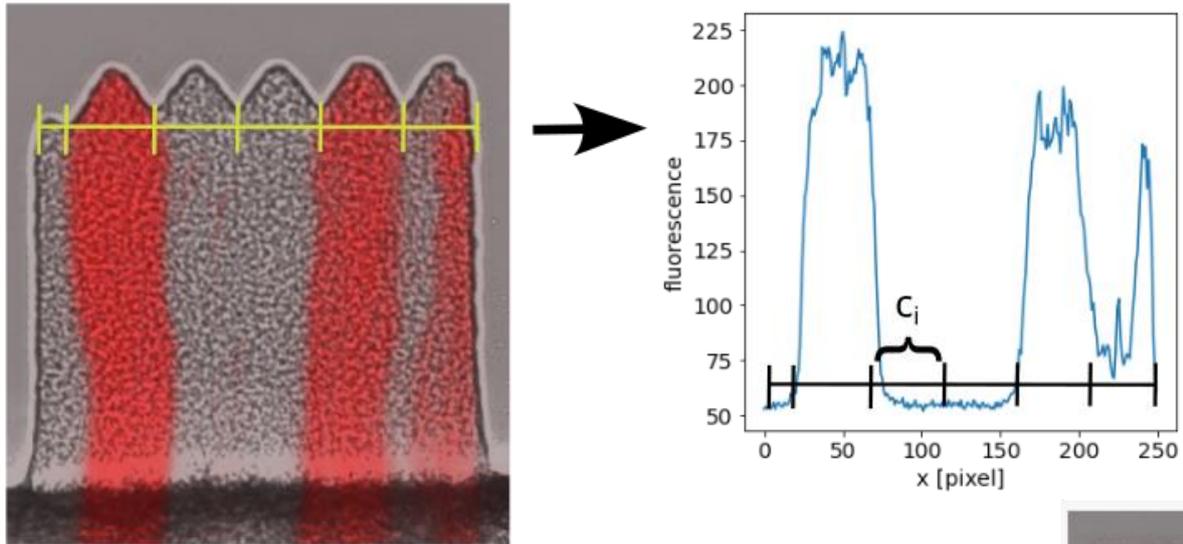

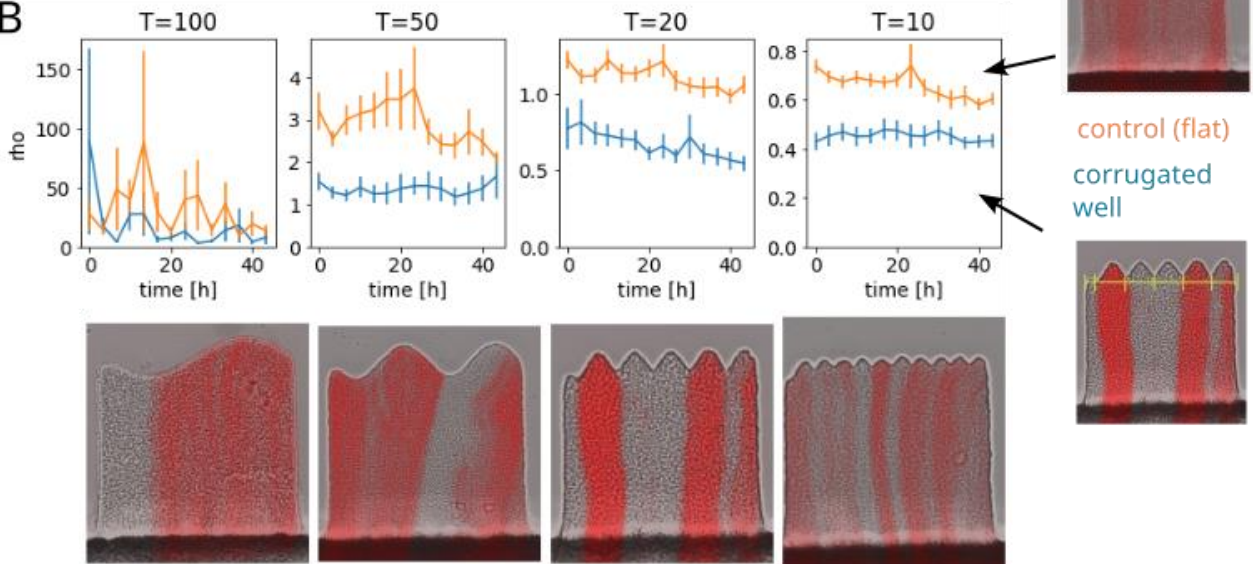

**Fig. S2. Intra- versus inter-well heterogeneity. (A)** Illustration of the method with which fluorescence heterogeneity is quantified. $c_i$ is the vector of pixel intensities in the $i$th "pocket", calculated along the line (yellow) just above the undulations. The diversity ratio $\rho = \mathrm{E}(\mathrm{D}(c_1), \ldots, \mathrm{D}(c_N))/\mathrm{D}(\mathrm{E}(c_1), \ldots, \mathrm{E}(c_N))$ where $\mathrm{E}$ and $\mathrm{D}$ denote mean and standard deviation, respectively. **(B)** Diversity ratio $\rho$ versus time for wells with different undulation period $T$.

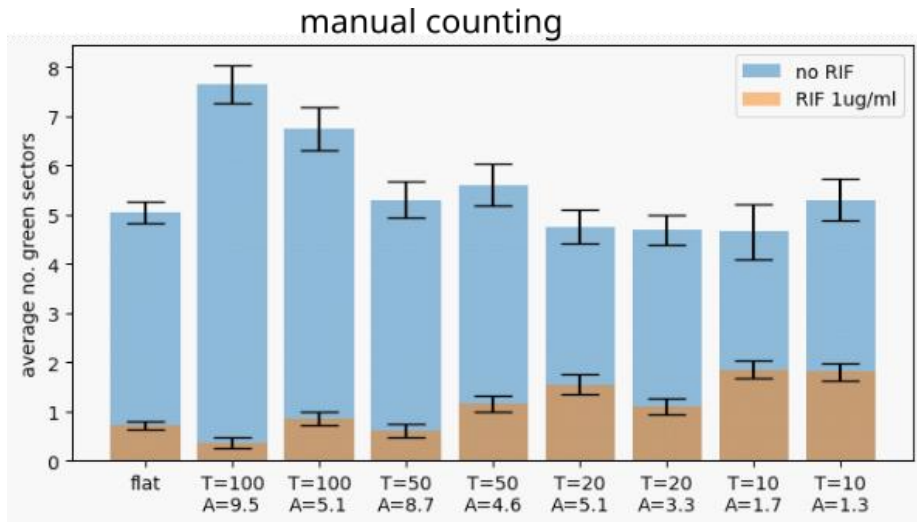

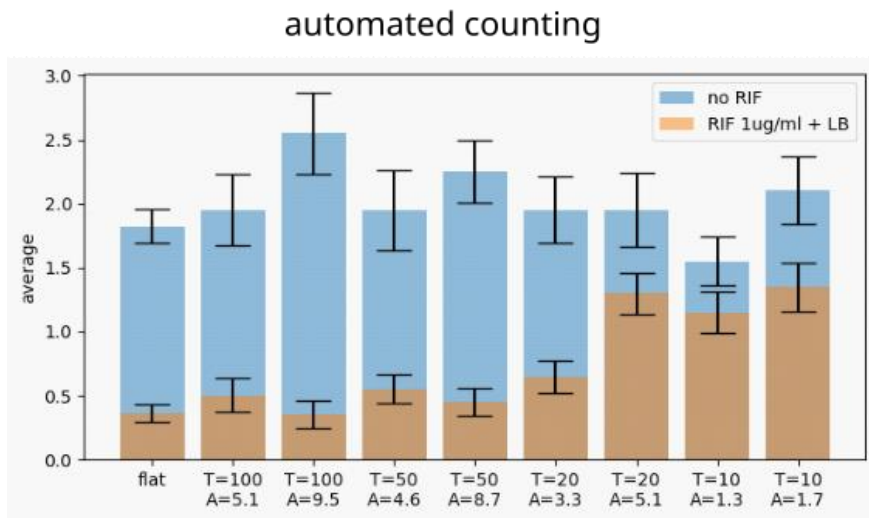

**Fig. S3. Average number of green sectors in the experiment from Fig. 4.** Upper panel: sectors counted manually, lower panel: sectors counted by a computer algorithm. The automated algorithm deliberately ignores very small sectors, thus the numbers are generally lower than with manual counting. However, both methods show the same trend: the ratio of the number of sectors after and before the RIF exposure increases with decreasing $T$.

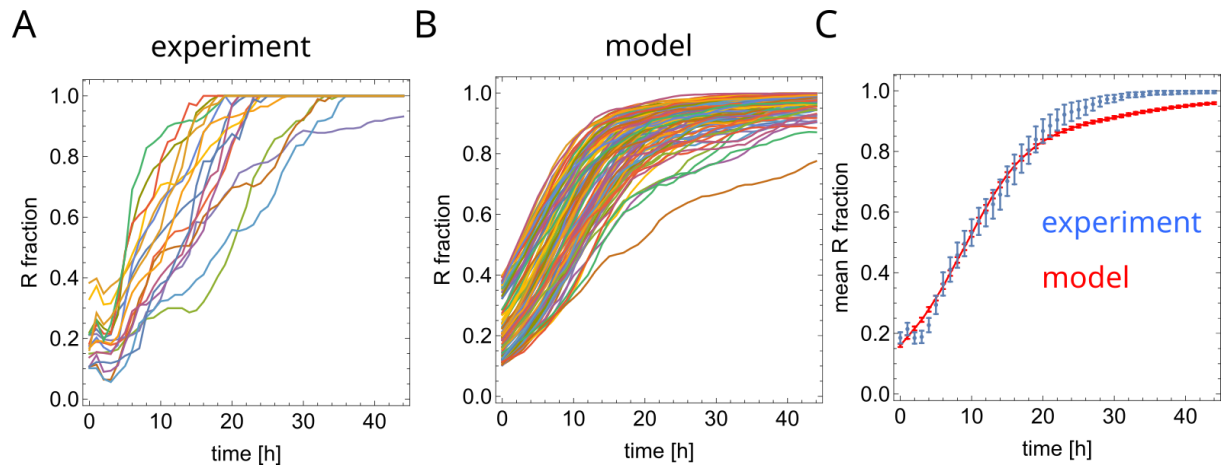

**Fig. S4. Relative fitness from sector expansion. (A)** The fraction of fitter (resistant) cells versus time in flat-bottomed wells (colors = different wells) during the 1 µg/ml RIF exposure experiment from Fig. 4C. Only wells in which the initial fraction at $t = 0$ h is between 0.1 and 0.4 have been selected. **(B)** The same fraction of fitter cells from the computer model, for the relative fitness of green (sensitive) to red (resistant) cells $W_{G/R} = 0.2$, initial fraction of fitter cells having the same distribution as in panel (A) and the doubling time of 3 h. **(C)** Mean fraction of resistant cells for the experiment and the model, for $W_{G/R} = 0.2$, which gives the best fit to the experimental data.

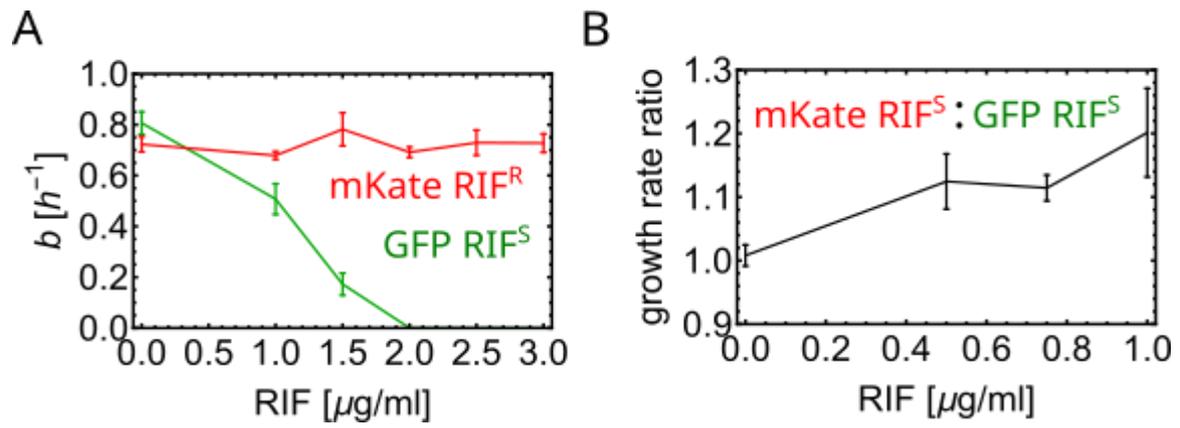

**Fig. S5**. **(A)** Growth rates of the RIF-sensitive GFP strain, and the RIF-resistant mKate strain. In the absence of RIF, the GFP strain grows ≈10% faster compared to the mKate strain. **(B)** The calculated ratio of the growth rates of mKate and GFP sensitive strains is consistent with no fitness difference in the absence of RIF, and a small growth advantage of mKate strain for non-zero RIF concentrations.

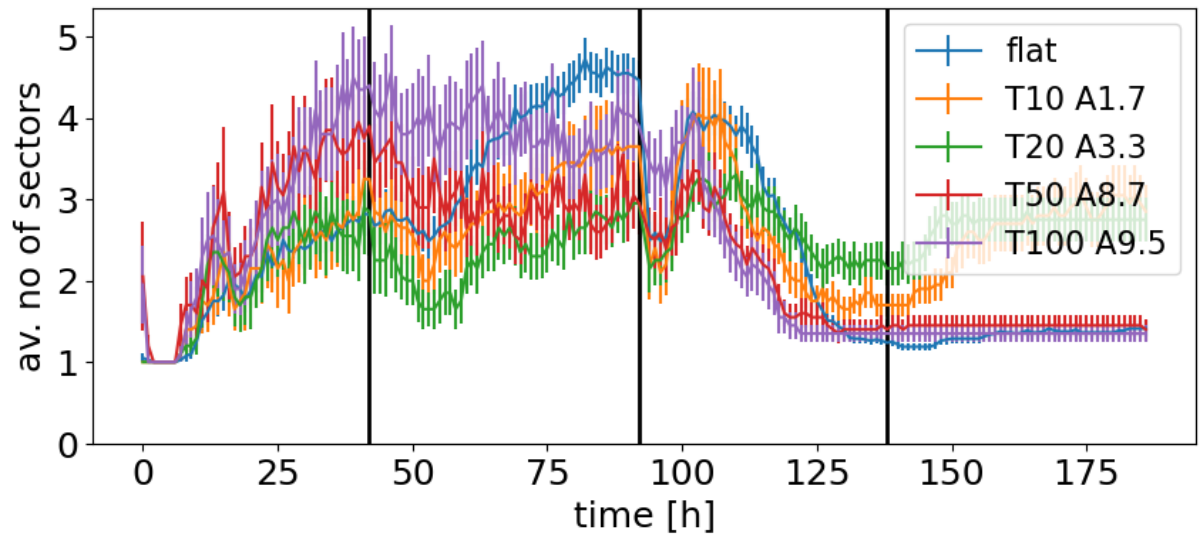

**Fig. S6. Average number of sectors (red + green) versus time, for the experiment from Fig. 4**. The sectors have been counted using the same automated algorithm as in Fig. 4.